\documentclass[%
 reprint,
 amsmath,amssymb,
 aps,
]{revtex4-2}

\usepackage{graphicx}
\usepackage{dcolumn}
\usepackage{bm}

\begin{document}

\preprint{APS/123-QED}

\title{Photoluminescence Features of Few-Layer Hexagonal $\alpha$-In$_2$Se$_3$}


\author{I.A. Eliseyev}%
 \altaffiliation[]{These authors contributed equally: I.A. Eliseyev, A.I. Veretennikov}
\author{A.I. Veretennikov}%
\altaffiliation[]{These authors contributed equally: I.A. Eliseyev, A.I. Veretennikov}
\author{A.I. Galimov}%
\author{L.V. Kotova}%
\author{G.V. Osochenko}%
\author{K.A. Gasnikova}%
\author{D.A. Kirilenko}%
\author{M.A. Yagovkina}%
\author{Yu.A. Salii}%
\author{V.Yu. Davydov}%
\author{P.A. Alekseev}%
\author{M.V. Rakhlin}%
\email{maximrakhlin@mail.ru}

\affiliation{Ioffe Institute, St. Petersburg, 194021, Russia}%




\date{\today}

\begin{abstract}
Indium (III) selenide is currently one of the most actively studied materials in the two-dimensional family due to its remarkable ferroelectric and optical properties. This study focuses on the luminescent properties of few-layer In$_2$Se$_3$ flakes with thicknesses ranging from 7 to 100 monolayers. To explore the photoluminescence features and correlate them with changes in crystal symmetry and surface potential, we employed a combination of techniques, including temperature-dependent micro-photoluminescence, time-resolved photoluminescence, Raman spectroscopy, atomic force microscopy, and Kelvin probe force microscopy. X-ray diffraction and Raman spectroscopy confirmed that the samples studied possess the $\alpha$-polytype structure. The micro-photoluminescence spectrum consists of two bands, A and B, with band B almost completely disappearing at room temperature. Temperature-dependent photoluminescence and time-resolved measurements helped us to elucidate the nature of the observed bands. We find that peak A is associated with emission from interband transitions in In$_2$Se$_3$, while peak B is attributed to defect-related emission. Additionally, the photoluminescence decay times of In$_2$Se$_3$ flakes with varying thicknesses were determined. No significant changes were observed in the decay components as the thickness increased from 7 to 100 monolayers, suggesting that there are no qualitative changes in the band structure.
\end{abstract}

\maketitle


\section{\label{sec:level1}Introduction}

The significance and diversity of two-dimensional (2D) layered materials with van der Waals bonding have grown substantially since the discovery of the quantum Hall effect in graphene in 2004 \cite{novoselov2004electric}. The emergence of graphene was followed by a surge of interest in 2D semiconductors, beginning with the discovery of bright photoluminescence in single-layer transition metal dichalcogenide (TMD) MoS$_2$ \cite{splendiani2010emerging} and the subsequent development of efficient transistors based on this material \cite{radisavljevic2011single}. Over the next decade, the family of 2D semiconductors expanded, and research gradually shifted from TMDs to other post-transition metal chalcogenides, such as InSe, GaSe, and In$_2$Se$_3$.

In$_2$Se$_3$ exists in various crystalline phases, with the layered $\alpha$- and $\beta$-phases being the most actively studied and utilized in device structures. Unlike TMDs, which are primarily valued for their optical and electronic properties, In$_2$Se$_3$ has shown promise for a broader range of applications. It has been recognized as both a ferroelectric \cite{huang2022two, mukherjee2022indium, xue2018room, cui2018intercorrelated} and piezoelectric \cite{xue2018multidirection, feng2016sensitive} material. Its potential applications include high-performance 2D strain sensors \cite{feng2016sensitive}, $\alpha$-$\beta$ transition-based phase change memory \cite{ignacio2024air}, and ferroelectric random access memory \cite{Miao_2023}.


Despite the considerable number of recent publications on the ferroelectric properties of $\alpha$-In$_2$Se$_3$, its optical characteristics, particularly in atomically thin layers, remain poorly understood. Several studies have investigated the photoluminescence (PL) of In$_2$Se$_3$ \cite{balakrishnan2018epitaxial}, as well as the layer-dependent variations in Raman and PL spectra for thicknesses ranging from 1 to 4 monolayers (ML) \cite{zhou2015controlled} for $\alpha$-In$_2$Se$_3$. However, regarding the nature of the PL, two studies examining the temperature dependence of In$_2$Se$_3$ PL \cite{balkanski1986photoluminescence, zhirko2018characterization} provide contradictory information about the origin of the main PL peaks.

Therefore, investigating the optical properties of In$_2$Se$_3$ is of significant importance. In this work, we study $\alpha$-In$_2$Se$_3$ flakes of the hexagonal 2$H$ polytype, with thicknesses ranging from 7 to 100 monolayers, which were obtained by micromechanical exfoliation and transferred onto an Au (50 nm)/Si substrate. The samples were characterized using temperature-dependent micro-photoluminescence ($\mu$-PL) spectroscopy, along with Raman spectroscopy, high-resolution transmission electron microscopy (HRTEM), atomic force microscopy (AFM), X-ray diffraction (XRD), and Kelvin probe force microscopy (KPFM).


\begin{figure*}[t]
    \includegraphics[width=1\textwidth]{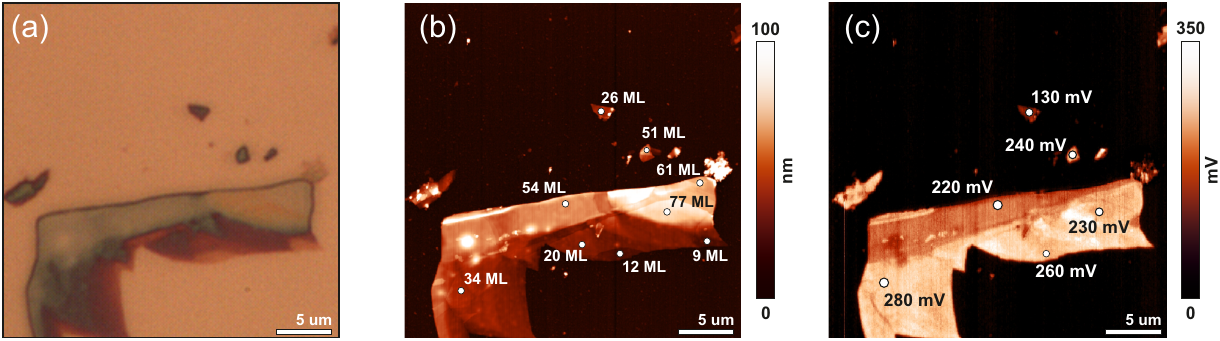}
    \caption{a) Optical image , b) topography  and c) surface potential distribution of one of the flakes under study. Flake thicknesses estimated from the AFM data are indicated on the topography map.}
    \label{KPFM}
\end{figure*}

\section{Samples}

Studied structures were fabricated from a commercially available high quality In$_2$Se$_3$ bulk crystal. Silicon wafers coated with 50-nm thick gold were used as substrates. For sample preparation, the so-called Scotch tape method was used (for more details see Appendix).

After fabrication, the topography of the samples was studied using AFM. Several flakes were selected based on their thickness. 
We picked flakes with thicknesses ranging from 7 to 100 MLs to investigate the evolution of optical properties of a few-layer In$_2$Se$_3$. The optical image of one of the flakes under investigation is shown in Fig. 1(a), while Fig. 1(b) displays the topography of the selected flake, along with the corresponding number of layers measured by AFM.


Figure 1(c) shows the surface potential distribution obtained by KPFM during simultaneous AFM scanning. KPFM measurements of van der Waals materials typically reveal a dependence of surface potential on the number of layers, attributed to variations in the work function \cite{Borodin_2019, Borodin2024}. As depicted in Figure 1, the flakes studied here are large enough to enable reliable data collection on both PL and surface potential. However, the ferroelectric properties of In$_2$Se$_3$ complicate the typical relationship between the work function and the number of layers due to the strong out-of-plane spontaneous polarization. Domains with different polarization orientations can be randomly distributed, resulting in surface potential changes on the order of hundreds of millivolts. \cite{Jia_2023}. In this study, we did not observe a consistent correlation between measured thicknesses and surface potentials, primarily due to the dominant influence of the electrical polarization direction distribution on the surface potential.

\begin{figure*}
            \includegraphics[width=\textwidth]{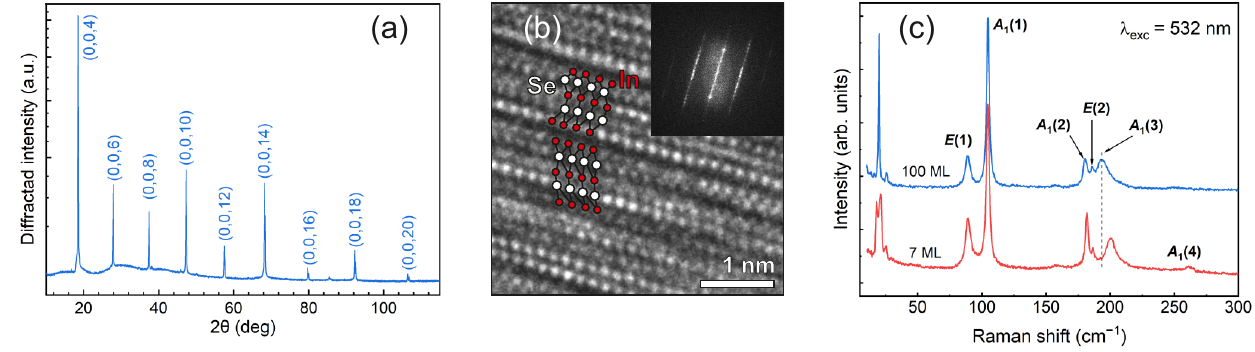}
            \caption{a)The XRD pattern of the bulk $\alpha$-In$_2$Se$_3$ crystal. b) HRTEM image of the bulk $\alpha$-In$_2$Se$_3$ crystal. Sketch shows a possible stacking variant for the 2H-polytype; the red and white circles represent the In and Se atoms, respectively. The inset show FFT image used for determining the lattice constant parameter. c) Raman spectra of the thinnest (7 ML) and thickest (100 ML) areas of the studied flakes. The spectra are normalized to the intensity of the $A_1(1)$ line and vertically shifted for clarity. The dashed line demonstrates the position of the $A_1(3)$ line in the spectrum of the 100-layered flake (192 cm$^{-1}$). }
            \label{XRD-TEM}
        \end{figure*}     

        \begin{figure} [t]
    \centering
    \includegraphics[width=1\linewidth]{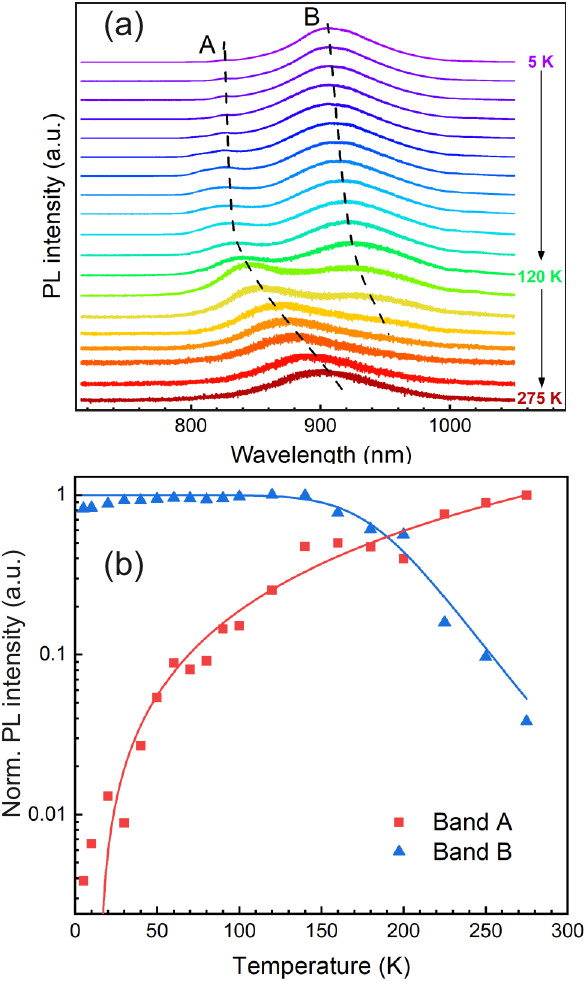}
    \caption{a) Normalized PL spectra (vertically shifted for clarity) measured in a bulk $\alpha$-In$_2$Se$_3$ at different temperatures with 532 nm $cw$ excitation. A and B denote the A and B bands, respectively. b) Temperature dependence of the normalized integrated PL intensity measured in a bulk $\alpha$-In$_2$Se$_3$.}
    \label{T-data}
\end{figure}

\begin{figure*}
    \flushleft
    \includegraphics[width=1\textwidth]{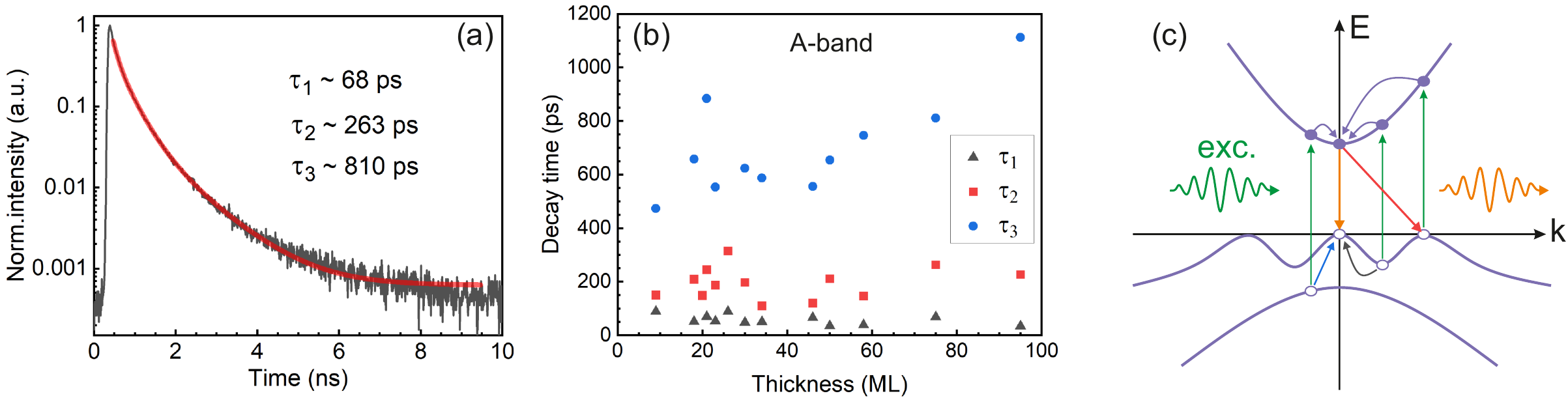}
    \caption{a) Normalized PL decay curve measured from 20-ML $\alpha$-In$_2$Se$_3$ flake. b) Decay times $\tau_1$, $\tau_2$ and $\tau_3$ (fast, medium and slow, respectively) as a function of the number of layers measured for the A-band. c) The scheme of excitation, relaxation, and emission processes
in a multilayer In$_2$Se$_3$ at high excitation energy. The orange and red arrows show direct and indirect optical transition, respectively. Green arrows illustrate the absorption of excitation photons; purple arrows illustrate electron relaxation on the bottom of the conduction band. Blue and gray arrows denote the hole relaxation paths corresponding to the "slow" and "fast" decay components.}
    \label{A_T-data}
\end{figure*}

\subsection{Crystal structure}

Figure \ref{XRD-TEM}a shows a typical XRD pattern of the bulk In$_2$Se$_3$ crystal used for exfoliation. The obtained peak positions confirm the $\alpha$-polytype of the sample and the hexagonal phase \cite{NGUYEN2019151968}.
A slight broadening can be caused by the presence of two-dimensional defects perpendicular to the stacking plane due to the van der Waals crystal structure of $\alpha$-In$_2$Se$_3$.


To determine the quality of the crystal structure, we conducted HRTEM studies. The results of the HRTEM studies of a bulk In$_2$Se$_3$ crystal are presented in Fig. 2b. The structure shows abundant stacking faults in the basal plane (001) that is typical for 2D crystals. The stacking faults are well seen in the HRTEM image as dark lines. The Fast Fourier Transform (FFT) data obtained from the tiny area of Fig. 2b allowed us to determine the lattice constant $c$ = 19.56 $\text{\AA}$ in the sample under study. The obtained value is consistent with previous studies for hexagonal $\alpha$-In$_2$Se$_3$  \cite{liu2019atomically, Ho_2021}.

\subsection{Raman data}

Raman spectroscopy was used to determine the type of crystal structure and to estimate the thickness and composition of the flakes. Figure \ref{XRD-TEM}c shows characteristic spectra measured from different parts of the flake. The observed set of lines corresponds to the $\alpha$-polytype In$_2$Se$_3$, and the relatively high intensity of the line at $\sim$ 90 cm$^{-1}$, denoted as $E(1)$, indicates that the crystal under study belongs to the $2H$-polytype \cite{liu2019atomically}. Line designations are given according to irreducible representations of the $P3m1$ group \cite{eliseyev2024raman}. In our recent work, it was shown that the frequency of the $A_1(3)$ line is sensitive to the number of layers \cite{eliseyev2024raman}. Figure \ref{XRD-TEM}c shows that its frequency in the thinnest parts of the flakes (red spectrum) corresponds to 7-layer In$_2$Se$_3$, while in the thickest parts (blue spectrum) it corresponds to $\ge 70$ ML.

The number of layers determined from the Raman spectra is consistent with the AFM data, which give 7 ML for the red spectrum and 100 ML for the blue one. The changes observed in the low-frequency region of the spectra (10–30 cm$^{-1}$) upon transitioning from the red to the blue spectrum further support the thickness estimates provided above. Specifically, as the thickness increases from 7 to 100 MLs, the two peaks at 18 and 21 cm$^{-1}$ merge into a single peak with a maximum at 20 cm$^{-1}$, accompanied by a small low-frequency shoulder. We did not identify any signs of oxidation or structural degradation. In particular, the line at 130 cm$^{-1}$ corresponding to indium oxide \cite{gan2013oxygen} was not observed, and the intensity of the peak at 252 cm$^{-1}$ corresponding to oxidation-associated Se bridge defects \cite{mukherjee2020scalable} was negligible compared to that of the $A_1(4)$ peak. Thus, the Raman data confirm the thickness of the flakes measured by AFM and indicate that the material is pristine $\alpha$-$2H$-In$_2$Se$_3$, with no significant oxidation.

\section{Emission properties}
\subsection{Micro-photoluminescence}
The $\mu$-PL spectrum of a bulk $\alpha$-In$_2$Se$_3$ consists of two main bands denoted as A and B for the high-energy and low-energy one, respectively. The data on the origin of these bands are quite contradictory. Thus, Zhirko et al. observed similar bands in the PL spectra of Bridgeman-grown bulk In$_2$Se$_3$ and associated the A-band with the luminescence of In$_2$Se$_3$ and the B-band - with the contribution of InSe nanocrystals formed in the In$_2$Se$_3$ matrix during growth \cite{zhirko2018characterization}. However, in some works the B-band is associated with the emission of defects, while the A-band -- with direct interband transitions \cite{balkanski1986photoluminescence, Kambas88disorder}. 

To investigate the origin of the spectral bands, we have measured the temperature dependence of the $\mu$-PL of bulk $\alpha$-In$_2$Se$_3$ (Fig. \ref{T-data}a). In the measured PL spectra, two bands A and B are observed, and at low temperatures the PL intensity of B-band is significantly greater than that of A-band. As the temperature increases, a redistribution of intensities occurs and at temperatures above 200 K B-band practically disappears. Figure \ref{T-data}b shows the normalized integrated intensity of each band as a function of temperature. The experimental data were fitted using the formula described by Mott \cite{mott1938absorption}:
\[f_i(T) = \frac{p_i(T)}{P_0(T)},\ \ p_i(T) = P_ie^{-W_i/kT},\ \ i=A,B,C,\]
where $k$ is the Boltzmann constant, $T$ is the temperature, and each of the $p_i(T)$ and $W_i$ defines the rate and activation energy of the corresponding electron-hole process, respectively. Thus, $p_C(T)$ describes nonradiative recombination of holes and electrons, whose energy transfers directly to heat, $p_A(T)$ describes the process of direct interband radiative recombination of electrons and holes, and $p_B(T)$ describes the radiative recombination from defect levels. The overall rate $P_0(T) = \sum_ip_i(T)$ is important to obtain the normalized probability of each contribution, and $P_i$ sets the the relative impact of the recombination rate $p_i(T)$ in the overall rate $P_0(T)$. 

At low temperatures, the electron-hole pairs excited by the laser quickly relax to defect levels in the In$_2$Se$_3$ energy band gap, and then recombine from them. Therefore, a high PL intensity can be observed in the B-band. As the temperature increases, electrons at the defect levels begin to thermalize (see blue triangles in Fig. \ref{T-data}b). As a result, the PL caused by defects is suppressed, and electrons recombine with holes directly from the conduction band of $\alpha$-In$_2$Se$_3$. The increasing number of the direct interband transitions leads to increasing PL intensity in the A-band (see red squares in Fig. \ref{T-data}b). Thus, the observed temperature dependence of the integrated PL indicates that the luminescence in the B-band is defect-related, while the PL in the A-band is associated with direct interband transitions.

\subsection{Time-resolved photoluminescence}
Previous studies of related layered materials have shown that as the number of layers decreases, the band structure changes significantly \cite{Borodin2024, Zeineb2017PRB, Yi2013NatNan}. To investigate the influence of the number of layers on the band structure in $\alpha$-In$_2$Se$_3$, we performed TRPL measurements for $\alpha$-In$_2$Se$_3$ flakes of different thicknesses. Since the PL spectrum contains two overlapping bands, TRPL measurements for each thickness were performed separately for a small spectral window ($\Delta\lambda\sim$ 5 nm) in the center of each band.


The typical TRPL data for the B-band exhibit exponential decay with characteristic lifetimes exceeding 100 ns for all thicknesses. These long decay times confirm the association of the B-band with defect-induced recombination, which is consistent with our interpretation based on the temperature dependence of the PL spectra and the large number of stacking faults obtained from the HRTEM measurements.

TRPL measurements of the A-band revealed that a satisfactory fit of the PL intensity decay could only be achieved using three independent exponential components. The typical decay curve measured from a 20-ML In$_2$Se$_3$ flake is shown in Fig. 4a. The decay times, $\tau_j$, for different thicknesses, obtained from the fit, are illustrated in Fig. \ref{A_T-data}b. As evident from the data, no significant changes in the decay components are observed with increasing thickness, suggesting the absence of qualitative changes in the band structure as the number of layers increases from 7 to 100 ML.

For Band A, three distinct decay components were identified: the "fast" component with a characteristic decay time $\tau_1 \sim 40$–-100 ps, the "medium" component with a decay time $\tau_2 \sim 100$-–250 ps, and the "slow" component with a decay time $\tau_3 \sim$ 0.47-–1.1 ns (see Fig. \ref{A_T-data}b). Since some of the measured decay times for the "fast" component fall below the detection limit of the setup (40 ps), the observed recombination may occur even faster than the time resolution of our equipment. We attribute the "fast" component to direct interband transitions of free carriers or bound excitons, as indicated by the respective hole relaxation process shown with the gray arrow in Fig. \ref{A_T-data}c. According to density functional theory calculations \cite{Li2018, li2020band}, the valence band extrema in layered $\alpha$-In$_2$Se$_3$ can occur not only at the $\Gamma$-point, but also in its close vicinity (Fig. \ref{A_T-data}c). In this case, recombination of an electron at the bottom of the conduction band ($k=0$) and a hole at a valence band extremum ($k \neq 0$) leads to the appearance of the "medium" component (red arrow in Fig. \ref{A_T-data}c). The "slow" component is likely attributed to interband relaxation of holes generated in the lower valence bands by the laser pulse \cite{Borodin2024} (blue arrow in Fig. \ref{A_T-data}c).

\section{Conclusions}

In conclusion, we have experimentally investigated the optical properties of few-layer $\alpha$-In$_2$Se$_3$ at cryogenic temperatures. Using the exfoliation technique, we prepared flakes with varying thicknesses in the range of 7 to 100 MLs. XRD measurements and Raman spectra of the bulk In$_2$Se$_3$ crystal used for exfoliation confirmed that the sample exhibits the $\alpha$-type crystal structure. HRTEM studies enabled us to determine the lattice constant and identify a significant number of stacking faults in the bulk material. Two distinct bands, A and B, were observed in the PL spectrum. From temperature-dependent PL and time-resolved PL measurements, the B-band was attributed to radiation related to defects, while the A-band was assigned to direct interband transitions.

For all studied thicknesses, the PL decay of the A-band was best fitted using three exponential components. The "fast" component has a decay time of $\tau_1 \sim 40$-–100 ps, the "medium" component decays with a time scale of $\tau_2 \sim 100$–-250 ps, and the "slow" component has a decay time of $\tau_3 \sim 0.47$–-1.1 ns. The "fast" component is attributed to the interband recombination of electrons and holes, while the "medium" component is associated with the recombination of an electron at the bottom of the conduction band ($k=0$) and a hole at the valence band extremum ($k \neq 0$). The "slow" component is likely due to the relaxation of holes generated in the lower valence bands by the laser pulse.

No significant correlations were found between decay times, layer thickness, and surface potential, indicating that the band structure remains largely unchanged across the $7–100$ ML range. Additionally, since the surface potential is influenced by domains with different spontaneous polarization orientations, the direction of polarization does not have a pronounced effect on the decay times.

We believe our findings provide important insights for the potential fabrication of In$_2$Se$_3$-based devices.

\begin{acknowledgments}
The work of M.V.R. was supported by a grant from the Russian Science Foundation (no. 24-72-00148, https://rscf.ru/project/24-72-00148/).
\end{acknowledgments}

\appendix

\section{Methods}
\subsection{Sample preparation}
The flakes were prepared using a mechanical micro-cleavage method using adhesive tape (Nitto ProTechno SPV-224). The method involves gluing the tape onto the surface of the material and peeling off the top layer. After peeling off, the tape with the upper part of the crystal stuck to it is folded in half and then released. This procedure is repeated 10-20 times, during which the flakes disintegrate into increasingly thinner ones. Eventually, very thin flakes are attached to the film and easily transferred to a suitable substrate. Using the layering technique described above, flakes of 7--100 ML in thickness were prepared, which were then transferred to a gold-coated substrate through a transfer polymer (Gel-Pak PF-X4) using a 2D crystal transfer system (HQ graphene). 

\subsection{AFM characterization}
The AFM study was carried out on an Ntegra Aura scanning probe microscope (NT-MDT, Russia). To obtain the topography of the samples, high-precision probes HA\_NC (NT-MDT, Russia) with a tip curvature radius of less than 10 nm, a resonant frequency of 235 kHz, and a force constant of 12 N/m were used. The surface potential was studied using conductive probes NSG10/Pt (NT-MDT, Russia) with a tip curvature radius of 35 nm, a resonant frequency of 81 kHz, and a force constant of 6 N/m.

\subsection{Raman and $\mu$-PL measurements}
Raman and PL measurements were performed using a Horiba LabRAM HREvo UV-VIS-NIR-Open spectrometer coupled with a closed-cycle hellium cryostat (CRYO Industries of America). The spectra were obtained in backscattering geometry with continuous wave (cw) excitation from the 532 nm laser line of a Nd:YAG laser (Laser Quantum Torus). To prevent damage to the flakes, the incident laser power was limited to 200 $\mu$W.

Raman measurements were carried out at room temperature. For them, an Olympus MPLN100$\times$ objective lens (NA = 0.9) was used, allowing us to obtain information from a region of diameter $\sim$ 1 $\mu$m. Raman spectra were acquired with a spectral resolution of 0.7 cm$^{-1}$ using a 1800 g/mm grating and a Symphony BIUV detector (Horiba). To suppress Rayleigh scattering and retrieve information from the ultra-low-frequency range (5--50 cm$^{-1}$), a set of Bragg filters (BragGrate) was used.

PL measurements were performed with the sample mounted into a cryostat, which allowed to control its temperature in the range of $5-275$ K. A long working distance objective (Leica PL FLUOTAR 50$\times$ L, NA = 0.55) was used to focus the laser beam and collect the PL signal from the sample, and the spectra were obtained using a 600 g/mm grating and a Synapse EMCCD detector (Horiba). 

\subsection{Time-resolved PL measurements}
The PL kinetics studies were performed at 8 K using a flow-through helium cryostat. PL excitation at 700 nm was performed by a picosecond pulsed Ti:sapphire laser with a pulse repetition rate of 80 MHz and a power density of 4 W/cm$^2$. For measuring the decay times in the B-band, a pulse repetition frequency of 1 MHz was chosen. Superconducting single-photon detectors (Scontel) with a time resolution of about 40 ps were used as TRPL detectors. Long-pass and short-pass tunable interference optical filters were used to measure decay curves in the maximum of the band. Spectral window was chosen $\sim$ 5 nm.

\nocite{*}

\bibliography{apssamp}

\end{document}